\documentclass[12pt]{spieman}  % 12pt font required by SPIE;
\usepackage{amsmath,amsfonts,amssymb}
\usepackage{graphicx}
\usepackage{setspace}
\usepackage{tocloft}
\usepackage{color,soul}
\usepackage[dvipsnames]{xcolor}
\usepackage{multicol}
\usepackage{siunitx}
\usepackage{booktabs}
\usepackage{nowidow}
\title{Benefits of adding radial phase dimples on scalar coronagraph phase masks}

\author[a*]{Niyati Desai}
\author[a]{Dimitri Mawet}
\author[b]{Eugene Serabyn}
\author[b]{Garreth Ruane}
\author[a]{Arielle Bertrou-Cantou}
\author[a]{Jorge Llop-Sayson}
\author[b]{A J Eldorado Riggs}

\affil[a]{California Institute of Technology, Pasadena, California, United States}
\affil[b]{Jet Propulsion Laboratory, California Institute of Technology, Pasadena, California, United States}

\cftpagenumbersoff{figure}
\cftpagenumbersoff{table} 
\begin{document} 
\maketitle

\begin{abstract}

Current scalar coronagraph focal plane mask designs are performance-limited by chromaticity. We investigate the effects of adding central Roddier and dual zone phase dimples to scalar vortex masks to improve broadband performance by suppressing the chromatic stellar leakage. We present hybrid designs with radial phase dimples integrated with the sawtooth vortex, wrapped vortex, and cosine phase mask. We show that using these dimples, it is possible to substantially improve the broadband contrast performance of scalar phase masks. We also show that although adding a phase dimple does not increase the sensitivity to low-order aberrations, suppressing the central leakage of scalar vortex coronagraphs does not restore the aberrations sensitivities to their notional state.
\end{abstract}

% Include a list of up to six keywords after the abstract

% Include email contact information for corresponding author
{\noindent \footnotesize\textbf{*}Address all correspondence to Niyati Desai,  \linkable{ndesai2@caltech.edu} }

% {\noindent \footnotesize{\textit{Paper received by JATIS Jan. 11, 2023; accepted for publication Mar. 24, 2023; published online Apr. 8, 2023.}}}

\begin{spacing}{1}   % use double spacing for rest of manuscript
% \begin{multicols}{2}

\section{INTRODUCTION}
\label{sec:intro}  % \label{} allows reference to this section

The recent decadal survey has established a primary goal of directly detecting and characterizing exoplanets for NASA's proposed flagship space telescope\cite{Astro2020_Report}. The direct imaging of exoplanets outside our solar system presents a significant challenge due to the small angular separation and substantial brightness contrast between stars and their planetary companions. Specifically, in the quest to directly identify Earth-like planets orbiting Sun-like stars, telescope instruments must be capable of capturing the reflected light from these planets, which is $10^{10}$ times fainter than that of their host stars in the UV, visible and near infrared spectral bands.

\begin{figure} [t]
\begin{center}
\includegraphics[height=4.7cm]{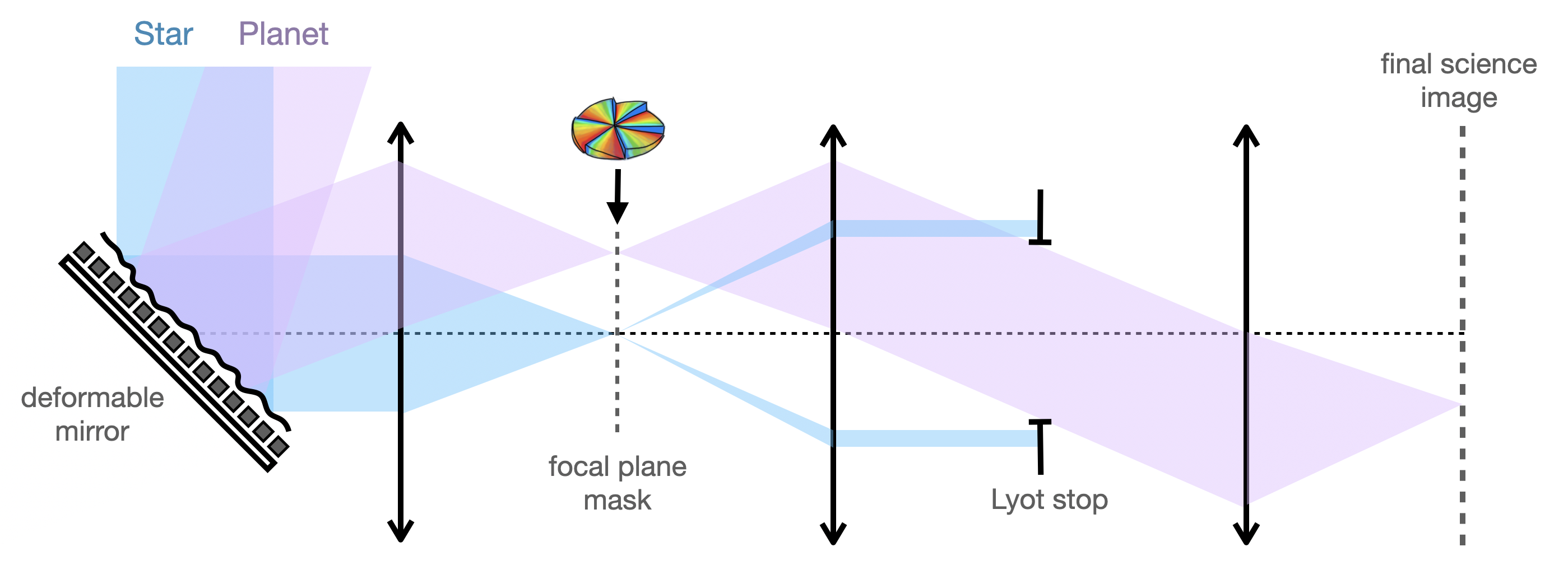}
\end{center}
\caption[fig:schema] 
%>>>> use \label inside caption to get Fig. number with \ref{}
{ \label{fig:schema} 
General schematic of a focal-plane mask based coronagraph. The focal plane mask shown here is a charge 6 scalar vortex phase mask.
}
\end{figure}

Coronagraphy is one approach to suppress on-axis starlight while allowing off-axis planet-light to pass through the telescope unaffected. The general schematic of a coronagraph is illustrated in Figure~\ref{fig:schema}. Coronagraphs typically use a focal plane mask (FPM), which, when precisely centered on a star, diffracts the starlight outside of a slightly undersized circular aperture in the subsequent pupil plane, known as the Lyot stop. Achieving coronagraphy at the $10^{-10}$ contrast level is exceptionally demanding and necessitates a precisely engineered and high-performance FPM. An especially crucial requirement is that the coronagraph operates in broadband light, so the atmospheres of detected planets can be spectrally characterized. Furthermore, this $10^{-10}$ contrast must be achieved at angular separations as close as 3.5 $\lambda/D$, where $\lambda$ is the wavelength and $D$ is the diameter of the primary mirror. This requirement comes from the median distance of HWO targets which is 1 astronomical unit (AU) at 14 parsecs (pc), or 1/14 arcseconds. The diffraction limit of a $D=6$ meter telescope at $\lambda = 550$ nm (V band) is $\lambda/D\times 206265=0.02"$. So a terrestrial planet with a separation of 1 AU at 14 pc observed by HWO would be found at 3.5 $\lambda/D$.

One of the primary challenges of coronagraphy is the presence of small optical aberrations, which can result in the leakage of starlight into the image plane. This results in speckles that are difficult to distinguish from actual planets. Modern coronagraphs now incorporate deformable mirrors (DMs) and closed-loop wavefront control to correct for these aberrations and create a speckle-free dark region, where exoplanets can be observed. Focal plane masks must ideally be insensitive to low-order aberrations to relax requirements on telescope and observatory stability.

The selection of a coronagraph for NASA's next flagship mission hinges on identifying a solution that achieves a contrast level of $10^{-10}$ at its inner working angle or within its dark hole region, maintains this performance within a 20\% bandwidth, exhibits insensitivity to select low-order aberrations, and ideally, is polarization-independent. As of now, no coronagraph has been able to demonstrate all of these characteristics in laboratory conditions.

% Currently one of the most promising coronagraphs is the optical vortex coronagraph\cite{Mawet2005, Foo2005}, however further design work and testing are needed to reach the milestone contrast performance.

In this paper, phase mask concepts introduced by Roddier \& Roddier (1997)\cite{Roddier_1997} and Soummer et al. (2003)\cite{Soummer2003} are revisited and investigated in combination with current scalar phase mask coronagraph concepts to suppress chromatic stellar leakage. This study investigates potential new designs which combine radial features with current azimuthal-only topographies using simulations of 2D wavefront propagation for these proposed scalar phase masks. The proposed mask designs would work in with an unobscured circular aperture, including segments (off-axis segmented which is presumably the baseline for HWO). More generally this work demonstrates the potential for design techniques  which could be applied to future phase mask designs to achromatize their performance. 

\subsection{Coronagraph Focal Plane Masks}
\label{subsec:fpms}

Starlight suppressing coronagraphic focal plane masks can be divided by their two working principles: amplitude masks and phase masks. The original Lyot coronagraph employs an amplitude mask consisting of an opaque central spot for the starlight to be blocked by and diffracted around. This occulting spot coronagraph still holds the record for the best recorded contrast with an unobscured, circular pupil: $4 \times 10^{-10}$ mean contrast with 10\% broadband light in a 360 degrees dark hole between 3 and 9 $\lambda/D$ \cite{Seo2019}. In pursuit of even better starlight rejection at closer separations, transmissive focal plane masks employing phase shifts were developed. This included the Roddier and Dual Zone phase mask, and then the four quadrant~\cite{Rouan_2000} and 8-octant phase mask~\cite{Murakami_2008}, finally leading to the development of the optical vortex phase mask~\cite{Mawet2005} along with the cosine phase masks~\cite{Henault2015}. The Roddier and Dual Zone phase masks use radially-varying phase features while the others use azimuthal-only topographies. This study focuses on combining both radial and azimuthal features in focal plane phase mask designs in an attempt to yield higher performance.

\subsection{Vortex Coronagraphs}
\label{subsec:vcs}

The vortex coronagraph uses a transparent phase-only mask in the focal plane which imparts a spiral phase shift, $e^{il\theta}$, on the incoming wavefront, where $\theta$ is the azimuth angle in the focal plane, and $l$ is the topological charge describing the number of spiral phase wraps the vortex mask imparts. Light from an on-axis point source, such as the star, that passes through the circular entrance pupil of radius $a$ is completely diffracted outside of the downstream Lyot stop of radius $b$, where $b \leq a$\cite{Mawet2005,Foo2005}. One of the key requirements that allows the vortex to produce theoretically perfect starlight rejection within the region of radius $b$, is that the topological charge must be an even nonzero integer\cite{Mawet2005}. The charge of a vortex trades off throughput at small angular separations against the sensitivity of the coronagraph to tip/tilt errors and other low-order aberrations. For future space telescopes, $l \geq 4$ will likely be desired to sufficiently relax low-order wavefront error requirement, however here only charge 6 vortex designs are considered since they were selected as baselines for the mission concepts the Habitable Exoplanet Imaging Mission (HabEx) and Large Ultraviolet Optical Infrared Surveyor (LUVOIR) \cite{Astro2020_Report,Ruane2018}.

Vortex coronagraphs come in two flavors: vector vortex coronagraphs (VVCs) and scalar vortex coronagraphs (SVCs). Both effectively imprint a spiral phase ramp onto the incoming wavefront, but differ in the fundamental mechanism they use to do so. Vector vortex masks use geometric phase shifts and are essentially half wave plates with a spatially varying fast axis. They imprint two different phase ramps ($e^{+il\theta}$ and $e^{-il\theta}$) on orthogonal circular polarization states. Current implementations of the VVC require an analyzer/polarizer to filter the chromatic leakage and also to isolate one sign phase ramp for downstream wavefront sensing and control. This is undesirable because it either effectively reduces the throughput by half or requires polarization splitting\cite{Ruane2020,Ruane2022}. In comparison, scalar vortex masks use longitudinal phase delays by varying the optical path difference introduced by the mask, which results in a chromatic behavior but no polarization dependence\cite{Ruane2019}. 

\subsection{Scalar Vortex Coronagraphs}
\label{subsec:svcs}

Scalar vortex masks have received much less attention than vector vortex masks, even though scalar vortex masks are polarization-insensitive and can theoretically provide sufficiently high throughput with simpler optics. The complex transmission of an SVC is highly chromatic: $ t(\theta,\lambda) = exp\big( i \frac{l_{0}\lambda_{0}}{\lambda} \theta \big) $, where $l_{0}$ is the designed vortex charge and $\lambda_{0}$ is the designed central wavelength. This effectively creates vortexes of non-integer charges for wavelengths offset from the central wavelength, which do not result in theoretically perfect starlight suppression \cite{Ruane2019}. A scalar vortex coronagraph with comparable capability to suppress starlight over a large bandwidth would therefore be advantageous over a VVC if it can also meet the stringent mission requirements for space coronagraphy: maintain high off-axis throughput (e.g., from a planet), have a small inner working angle, and be insensitive to tip/tilt errors and other low-order aberrations.

New designs are needed to make scalar vortex coronagraphs more achromatic and comparable in performance to vector vortex coronagraphs. Recently, there have been several different attempts to achromatize SVC designs including varying the FPM surface topography,\cite{Galicher2020, Desai_2022} combining two materials of different refractive indices into a dual-layer FPM,\cite{Swartzlander2006,Ruane2019} or employing metasurfaces optimized for broadband phase control\cite{konig2023, Palatnick23}.

% \subsection{Modal Decomposition}
% \label{subsec:modaldecomp}

\subsection{Radial Phase Mask Dimples}
\label{subsec:dimples}

This paper considers two specific radial phase mask designs: the Roddier \& Roddier dimple\cite{Roddier_1997} (referred hereafter as the `Roddier dimple') and the Dual Zone Phase Mask\cite{Soummer2003} (referred hereafter as the `DZPM dimple').

\begin{figure} [t]
\begin{center}
\includegraphics[height=5cm]{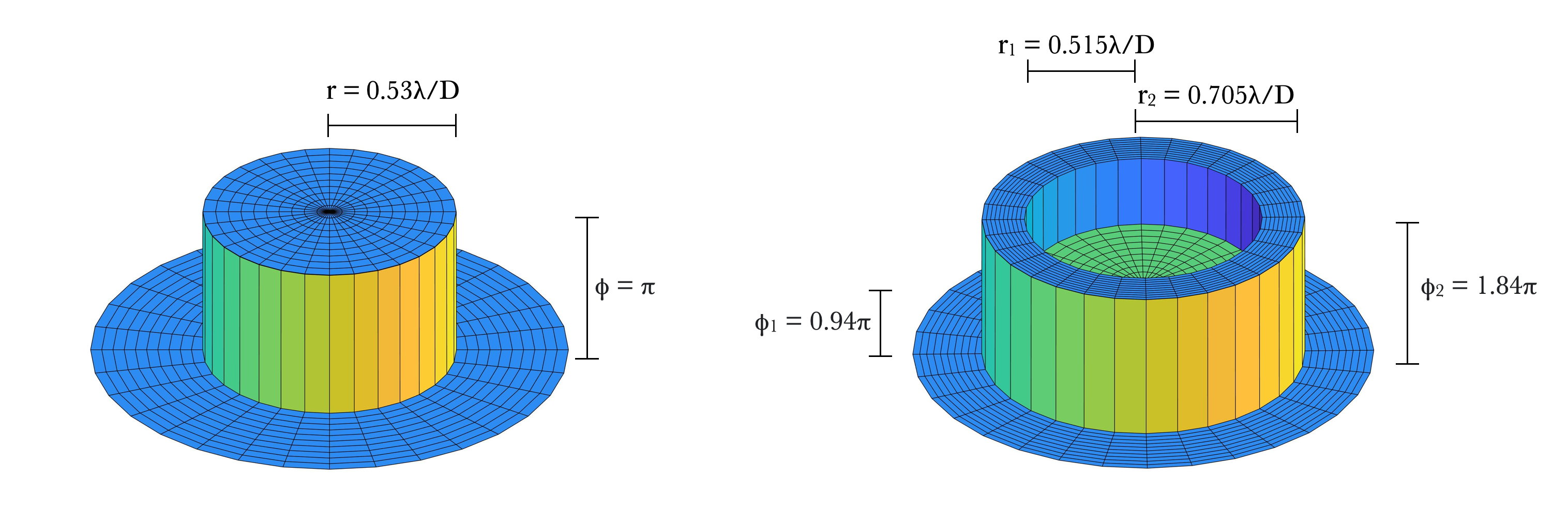}
\end{center}
\caption[fig:phasedimples] 
%>>>> use \label inside caption to get Fig. number with \ref{}
{ \label{fig:phasedimples} 
Left: Schematic with lateral dimensions and phase shift of the Roddier phase dimple\cite{Roddier_1997}. Right: Schematic with dimensions of the Dual Zone Phase Mask dimple\cite{Soummer2003}.
}
\end{figure}

Roddier \& Roddier (1997) originally proposed an improvement of Lyot's stellar coronagraph, by replacing the occulting mask with a transparent $\pi$ phase shifting mask\cite{Roddier_1997}. The left schematic in Figure~\ref{fig:phasedimples} shows the Roddier dimple dimensions. For the central wavelength $\lambda_{0}$, the phase shift of the mask should be $\pi$ and the size of the dimple should have a radius of $0.53 \lambda_{0}/D$. The optimal size for the Roddier dimple was found to be 0.43 times the size of the first dark ring in the Airy pattern since it encircles 50\% of the energy of the Airy pattern\cite{BornWolf1980}. The working principle is to create a phase-shifted region of a specific size so the exactly balanced fluxes, inside the dimple region and outside the dimple region, will cancel each other due to their opposite phases. There are two sources of chromatism in the dimple: the first is in the effective phase shift of the dimple, and the second is in the amount of flux going through the dimple. When the wavelength decreases slightly, the Airy disk shrinks so that more energy is concentrated in the central zone. 

Soummer et al. (2003) identified that the Roddier \& Roddier Phase Mask suffers from two chromatic effects: size and phase chromaticity. They proposed a generalization of the Roddier dimple, an alternative dual zone phase mask with better broadband performance\cite{Soummer2003} and conducted a more thorough optimization study of the parameter space for the size and phase shift of both zones. They also considered optimizing the DZPM for 20\%, 30\% and 40\% broadband coverage. One of their findings was that with large phase steps, the phase shift is more chromatic, making it more difficult to compensate over large bandwidths. Although the study recommended several DZPM solutions, most involved the use of an apodized pupil. The final parameters of the DZPM for an unapodized pupil are shown in the schematic on the right in Figure~\ref{fig:phasedimples}. For the central wavelength $\lambda_{0}$, the DZPM dimple consists of a central zone of radius $0.515 \lambda_{0}/D$ with a phase shift of $0.94\pi$ radians next to an annulus region with an outer radius of $0.705\lambda_{0}/D$ with a phase shift of $1.84\pi$ radians. Note, the performance of both the Roddier and DZPM coronagraphs is limited in terms of contrast for simple circular unobstructed apertures. However, although not considered in this study, solutions with apodized entrance pupils have shown to increase the contrast performance of both of these coronagraph masks\cite{Soummer2003}.

\section{HYBRID SCALAR PHASE MASK TOPOGRAPHIES}
\label{sec:topo}
We simulated several different azimuthally modulated scalar phase mask topographies of charge 6 to perform a direct comparison of their broadband performance. The Fast Linearized Coronagraphic Optimizer (FALCO)\footnote{\url{https://github.com/ajeldorado/falco-matlab}} software package was used to assess the chromaticity of the designs proposed in this paper as coronagraphic FPMs~\cite{Riggs2018}. The point spread function (PSF) of a star, and 2D wavefront propagation through all subsequent pupils and optics including the FPM was simulated to measure the contrast performance without any DM-assisted wavefront correction. To simulate the various FPMs proposed below, high resolution phase mappings were developed and resolution parameters were adjusted in FALCO's wavefront propagation to find an appropriate simulation environment for coronagraph mask development. In this study, four base scalar phase mask designs are considered:

\textbf{Classic Vortex:}
The first scalar phase mask design considered is the classic vortex. A classic scalar vortex is a simple phase ramp wrapping around the optical axis, and its slope is proportional to the charge of the vortex. For example a charge 6 classic vortex is a linear ramp from $0$ to $12\pi$, which in practice would result in thick optics less suited for the focal plane.

\textbf{Sawtooth Vortex:}
The second scalar phase mask design considered in this study is a phase-wrapped version of the classic vortex. It is referred to as the `sawtooth' vortex SVC because the azimuthal profile of this mask looks like a sawtooth pattern. In the sawtooth vortex design, $l$ phase ramps range from 0 to $2\pi$. The charge of the vortex, $l$, dictates the number of phase ramps, and therefore the slope (which is piece-wise identical to the classical case).

\textbf{Cosine:}
The third scalar phase mask design considered is the sinusoidal phase mask. This mask design is described by an azimuthal cosine-modulated phase function which spans 6 periods~\cite{Henault2015, Henault2016}.

\textbf{Wrapped Vortex:}
The fourth scalar phase mask design is the wrapped vortex which is inspired by the phase wrapping method proposed by Galicher et al. 2020~\cite{Galicher2020}. The design considered here is a charge 6 vortex version, introduced in Desai et al. 2022\cite{Desai_2022}. This design was obtained with an azimuthal profile optimization using the position of the 2$\pi$ jumps as parameters.

\begin{figure} [h]
    \begin{center}
    \begin{tabular}{c} %% tabular useful for creating an array of images 
    \includegraphics[height=7cm]{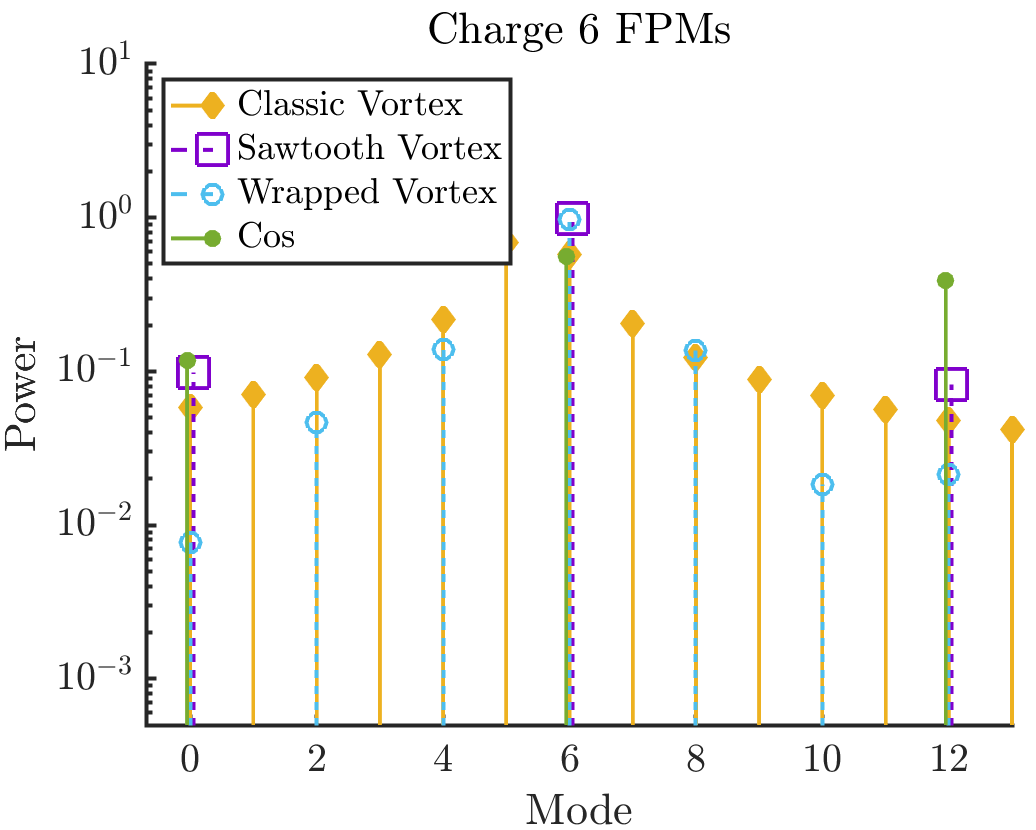}
    \end{tabular}
    \end{center}
    \caption[fig:moddecomp] 
    %>>>> use \label inside caption to get Fig. number with \ref{}
    { \label{fig:moddecomp} 
    Modal decomposition at a 10\% wavelength offset ($\lambda_{0}/\lambda = 1.1$) of the classical vortex (yellow), sawtooth vortex (purple), wrapped vortex (blue), and cosine phase mask (green).
 }
    \end{figure} 

To understand the chromaticity of these scalar phase masks, modal decompositions of the azimuthal phase profile for the four cases considered were performed at a 10\% wavelength offset ($\lambda_{0}/\lambda = 1.1$) and are shown in Figure~\ref{fig:moddecomp}~\cite{Ruane2018,Desai_2022}. This involves decomposing the azimuthal-only phase masks into an infinite sum of vortexes of integer charges. The complex transmission of any optical vortex can be written as:
\begin{equation}
\label{eq:fourier}
    t(\theta,\lambda)=\sum_{m} C_{m}(\lambda)e^{im\theta}
\end{equation}

with

\begin{equation}
\label{eq:coeffs}
    C_{m}(\lambda) = \frac{1}{2\pi} \int_{\pi}^{-\pi} t(\theta,\lambda) e^{-im\theta} d\theta
\end{equation}

And each mode \textit{m} corresponds to a charge m vortex whose behavior is known. Any odd or low even values less than the designed charge hurt performance via on-axis light leakage.

The modal decomposition of the sawtooth vortex reveals that no odd or low even modes emerge apart from the 0th order mode. The modes that emerge are at a base frequency of the charge, in this case 6 so there are modes at 6, 12, 18, etc. as seen in Figure~\ref{fig:moddecomp}. This property of the sawtooth offers a significant advantage over the classic vortex (for which the base frequency is 1), not only for increased insensitivity to low order wavefront aberrations, but also for improved broadband performance. (For further explanation of the modal decomposition analysis of SVCs, see Desai et al. 2022)\cite{Desai_2022}.

A modal decomposition for the sawtooth vortex and cosine phase mask reveals that the 0th order mode (central leakage) needs to be suppressed for better broadband performance. To suppress the 0th mode leakage, we explore adding central phase dimples. This is because central phase dimples only have 0th order leakage which can balance/subtract that of the vortex, without affecting higher azimuthal orders. This radial modulation is not expected to change the power distribution in the modal decomposition since it is only dependent on azimuthal features.

\begin{figure} [b!]
\begin{center}
\begin{tabular}{c} %% tabular useful for creating an array of images 
\includegraphics[height=3cm]{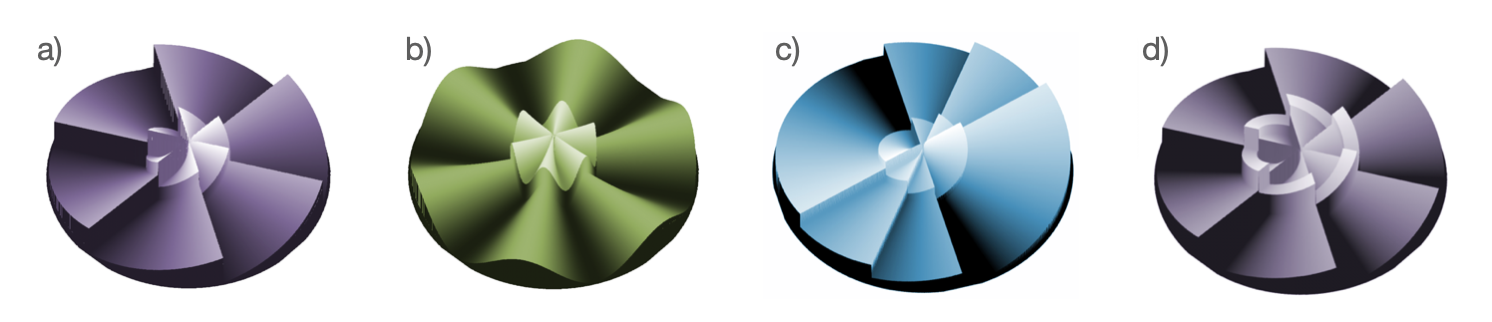}
\end{tabular}
\end{center}
\caption[fig:3d] 
%>>>> use \label inside caption to get Fig. number with \ref{}
{ \label{fig:3d} 
The 3D renderings of each of the focal plane phase masks considered in this study: (a) Sawtooth Vortex+Roddier, (b) Cosine+Roddier, (c) Wrapped Vortex+Roddier, (d) Sawtooth+Dual Zone Phase Dimple. }
\end{figure} 

This study investigates the effects of employing radial phase dimples in current scalar phase mask designs and the masks considered here add a Roddier dimple to the sawtooth vortex, wrapped vortex and cosine designs. Additionally, a DZPM dimple was added to the sawtooth vortex to investigate whether a significant effect on broadband performance resulted. Figure~\ref{fig:3d} shows the 3D renderings of these four focal plane scalar phase masks.

\section{RESULTS}
\label{sec:results}

\subsection{Contrast Performance}
\label{subsec:nowfc}

\begin{figure} [t]
\begin{center}
\includegraphics[height=9.5cm]{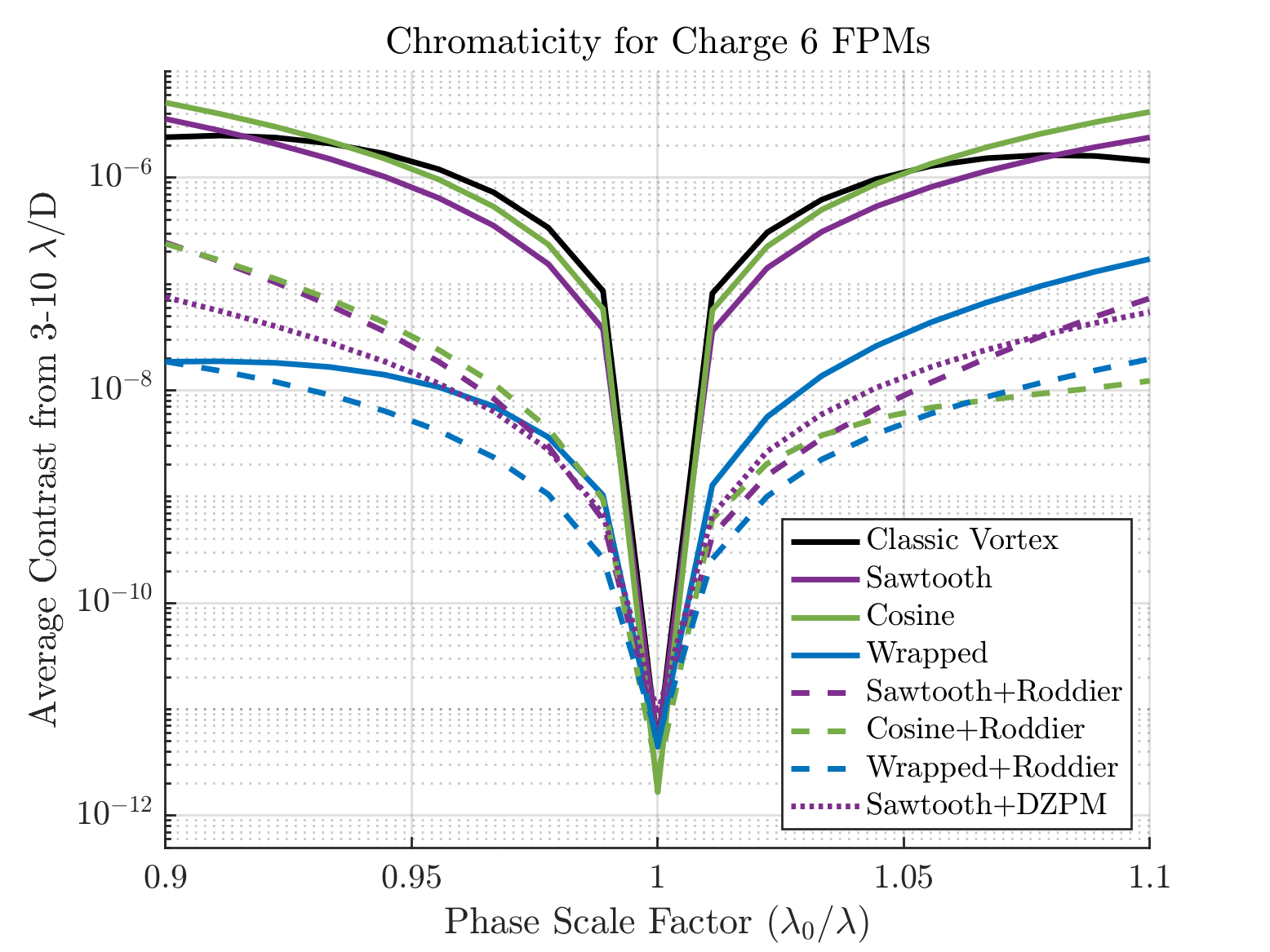}
\end{center}
\caption[fig:contrasts] 
%>>>> use \label inside caption to get Fig. number with \ref{}
{ \label{fig:contrasts} 
Chromaticity V-curve showing simulated average contrast from 3-10 $\lambda_{0}/D$ across a 20\% bandwidth. The solid lines correspond to the 4 base scalar phase mask designs considered in this study: classic vortex (black), sawtooth vortex (purple), cosine (green) and wrapped vortex (blue). The dashed lines correspond to the same 4 designs with a Roddier dimple added. The Dual Zone Phase Mask dimple with a sawtooth vortex (purple dashed) is also shown.}
\end{figure}

Each of the designs discussed hereafter in this paper was simulated and evaluated with 2D wavefront propagation. The average contrast in the 3-10$\lambda_{0}/D$ region was calculated for 19 discrete wavelengths equally spaced across a 20\% bandwidth. For 10\% bandwidth contrast measurements, the same spectral sampling was used. The raw contrast here is defined as the ratio of the average intensity in the region in the coronagraphic image to the highest intensity in the non-coronagraphic PSF. The resulting chromaticity V-curve in Figure~\ref{fig:contrasts} shows that although adding the Roddier dimple or DZPM to the sawtooth is still chromatic, it improves performance by at least an order of magnitude consistently across a 20\% bandwidth. More specifically, compared to a classic vortex, the contrast is improved by over 100× across a 10\% bandwidth ($\lambda_{0}/\lambda = 0.95$ or $\lambda_{0}/\lambda = 1.05$) and roughly 30× at the far edges of the 20\% bandwidth ($\lambda_{0}/\lambda = 0.9$ or $\lambda_{0}/\lambda = 1.1$).

\begin{figure} [h]
\begin{center}
\includegraphics[height=8cm]{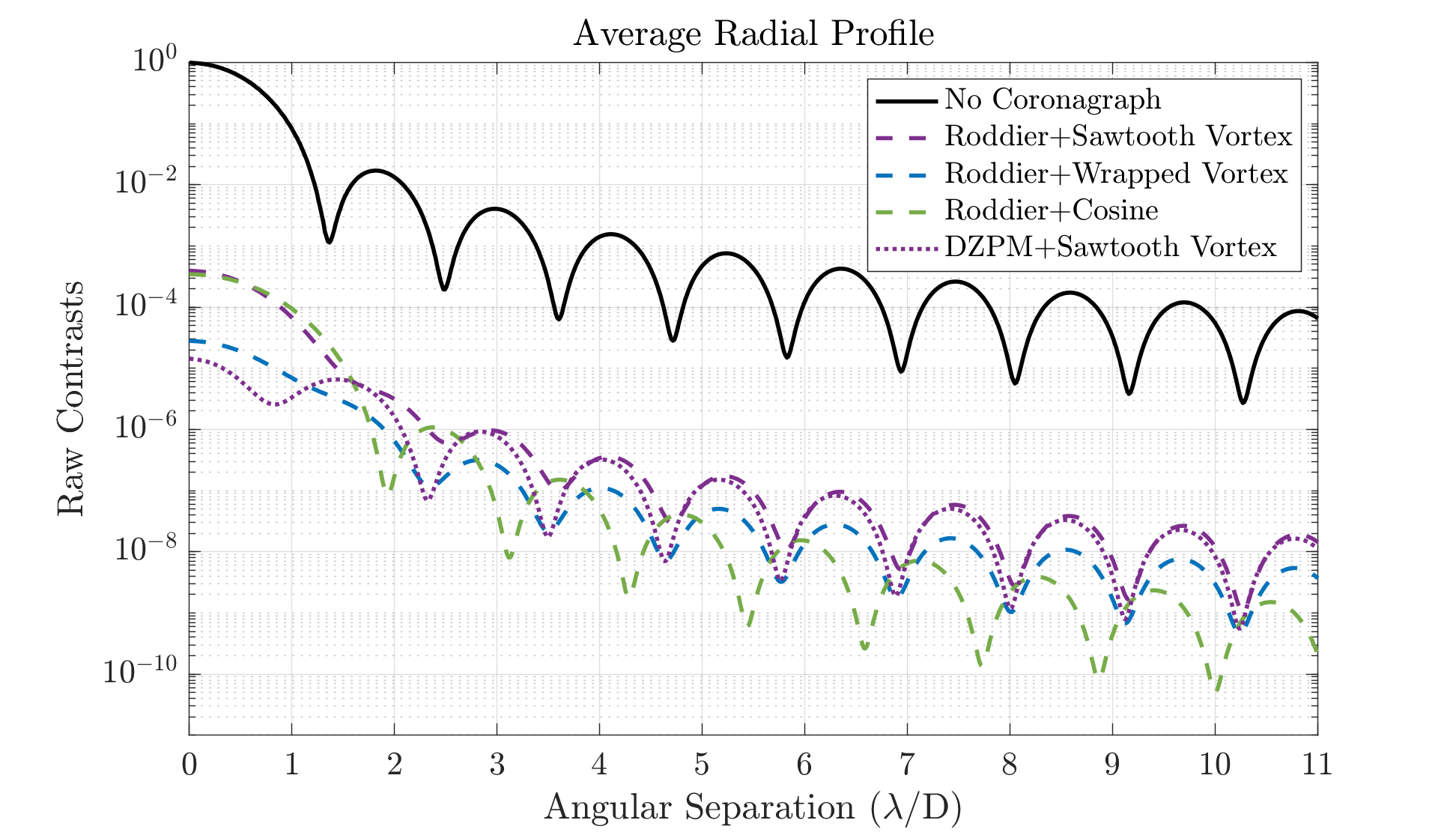}
\end{center}
\caption[fig:profs] 
%>>>> use \label inside caption to get Fig. number with \ref{}
{ \label{fig:profs} 
Average radial profiles for 20\% broadband light simulated for four proposed focal plane phase masks compared to the non-coronagraphic point-spread function (black). The other four profiles correspond to the coronagraphic focal plane image with the Roddier dimple added to the Sawtooth (purple dashed), Wrapped Vortex (blue dashed) and Cosine (green dashed) as well as the Dual Zone Phase Dimple added to the Sawtooth (purple dotted).}
\end{figure} 

The same simulation pipeline was used to compute the throughput of a simulated planet at angular separations ranging from 0 - 20 $\lambda_{0}/D$ for comparison between the sawtooth vortex mask and the hybrid vortex mask design. The addition of the central phase dimples was not expected to affect the throughput or inner working angle of a standard charge 6 vortex coronagraph. The simulation found the throughput as a function of angular separation for the sawtooth SVC and the Roddier+sawtooth SVC to be overlapping, both yielding 50\% throughput at an inner working angle of 3 $\lambda_{0}/D$.

Figure~\ref{fig:profs} compares the radial profiles of the 20\% broadband final coronagraphic image for each of the four proposed designs with the non-coronagraphic PSF (black). The figure illustrates the starlight attenuation effects of the Roddier dimple and the DZPM dimple and especially shows the contrast improvement compared to the non-coronagraphic profile.

Table~\ref{tab:contrasts} shows the average raw contrasts in 10\% and 20\% broadband light from 3-10 $\lambda_{0}/D$ in the simulated coronagraphic image for the four base scalar phase mask designs, the three hybrid Roddier dimple designs as well as the hybrid dual zone phase dimple design considered in this study. The 10\% broadband results show a factor of 2× contrast gained from using a sawtooth pattern instead of the classic vortex SVC. Furthermore, it can be clearly seen that adding a radial phase dimple in addition to using the sawtooth pattern provides a significant 108× factor improved average contrast over the classic vortex design. For 20\% broadband, the Roddier dimple provides a factor of 26× contrast improvement and the DZPM dimple offers a 49× factor of improvement over the sawtooth vortex alone. The DZPM dimple solution does not provide a significant advantage over the simpler Roddier dimple in 10\% bandwidth.

In Figure~\ref{fig:contrasts} the Roddier dimple not only shows a significant improvement when added to the sawtooth vortex design, but also for the cosine design. Table~\ref{tab:contrasts} shows a contrast improvement by a factor of 66× in 10\% and 46× in 20\% bandwidth. However for the wrapped vortex, adding the Roddier dimple only provides a factor of 4× for 10\% or 20\%. These results agree with the modal decompositions shown in Section~\ref{sec:topo}. Since the wrapped vortex already has a small 0th order mode, there is only a minor improvement to the central leakage from the Roddier dimple. 

{\renewcommand{\arraystretch}{1.2}
\begin{table}[!htbp]
    \centering
    \begin{tabular}{ccc}
    \toprule
         & \multicolumn{2}{c}{Bandwidth} \\
         & 10\% & 20\%\\
         \midrule
         % \cmidrule{2-3}
        Vortex & \num{4.06e-7} & \num{1.17e-6}\\ 
        Sawtooth & \num{2.03e-7} & \num{1.00e-6}\\
        Cosine & \num{3.16 e-7} & \num{1.56 e-6}\\
        Wrapped & \num{6.36 e-9} & \num{3.16 e-8}\\ \midrule
        Sawtooth + Roddier & \num{3.77 e-9} & \num{3.79 e-8}\\ 
        Cosine + Roddier & \num{4.76 e-9} & \num{3.30 e-8}\\ 
        Wrapped + Roddier & \num{1.40 e-9} & \num{6.62 e-9}\\ \midrule 
        Sawtooth + DZPM & \num{3.78 e-9} & \num{2.03 e-8}\\
    \bottomrule
    \end{tabular}

    \vspace{4 mm} 
    \caption{\label{tab:contrasts} Simulated broadband average raw contrast in 3-10 $\lambda_{0}/D$} 

\end{table}}

\subsection{Sensitivity to Low-Order Aberrations}
\label{subsec:sens}

\begin{figure} [t]
\begin{center}
\includegraphics[width=\columnwidth]{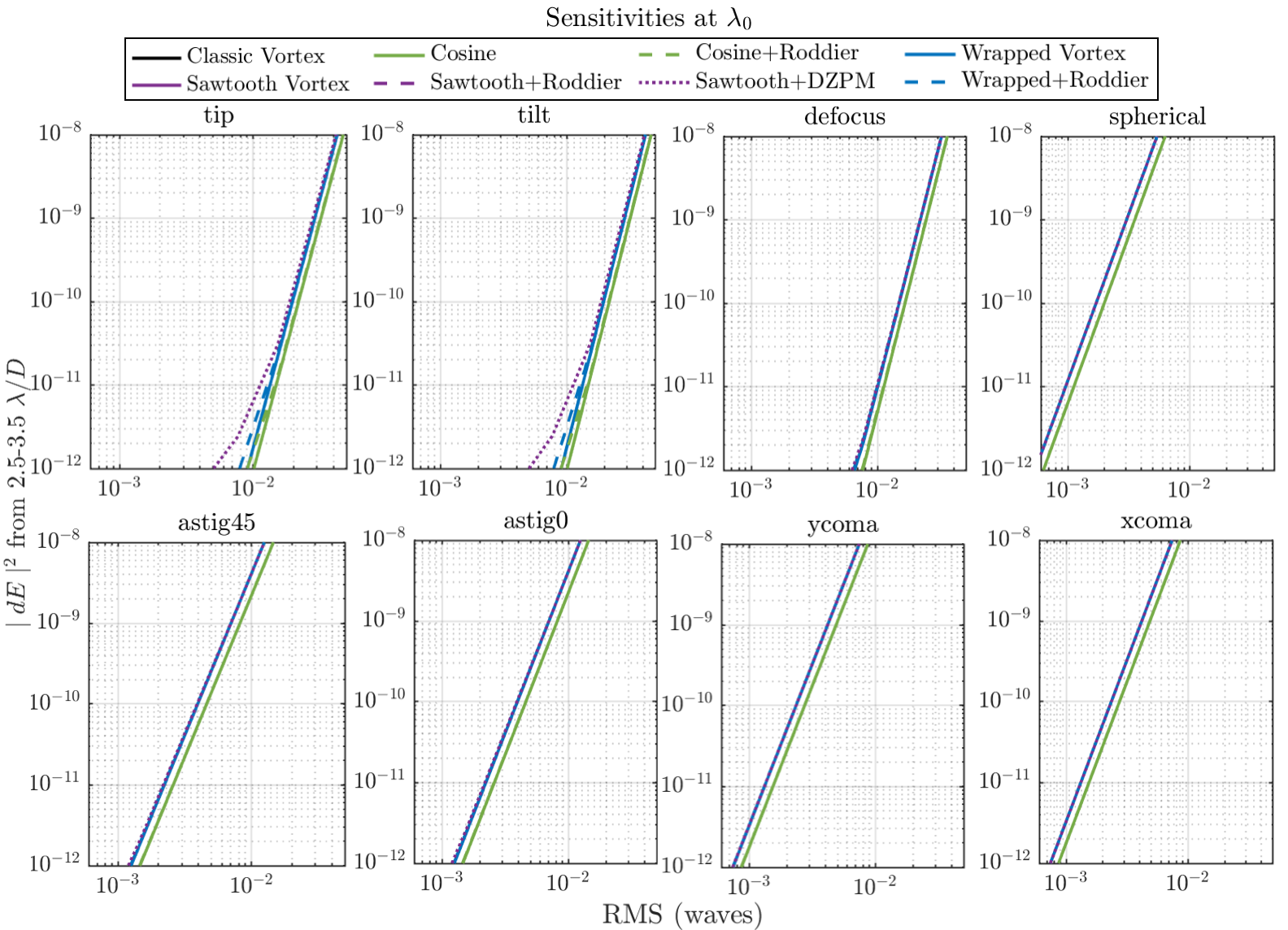}
\end{center}
\caption[fig:sens1] 
%>>>> use \label inside caption to get Fig. number with \ref{}
{ \label{fig:sens1} 
Monochromatic sensitivities to the lowest order Zernike aberrations (tip, tilt, defocus, spherical, astig, coma). The average change in stellar leakage across 3 ± 0.5 $\lambda_{0}/D$ is shown as a function of root-mean-square (RMS) wavefront error. The solid lines shown here are the four base focal plane masks (classic vortex, sawtooth vortex, wrapped vortex, and cosine). The dashed and dotted lines correspond to hybrid designs incorporating a Roddier phase dimple and dual zone phase dimple, respectively.}
\end{figure} 

\begin{figure} [t]
\begin{center}
\includegraphics[width=\columnwidth]{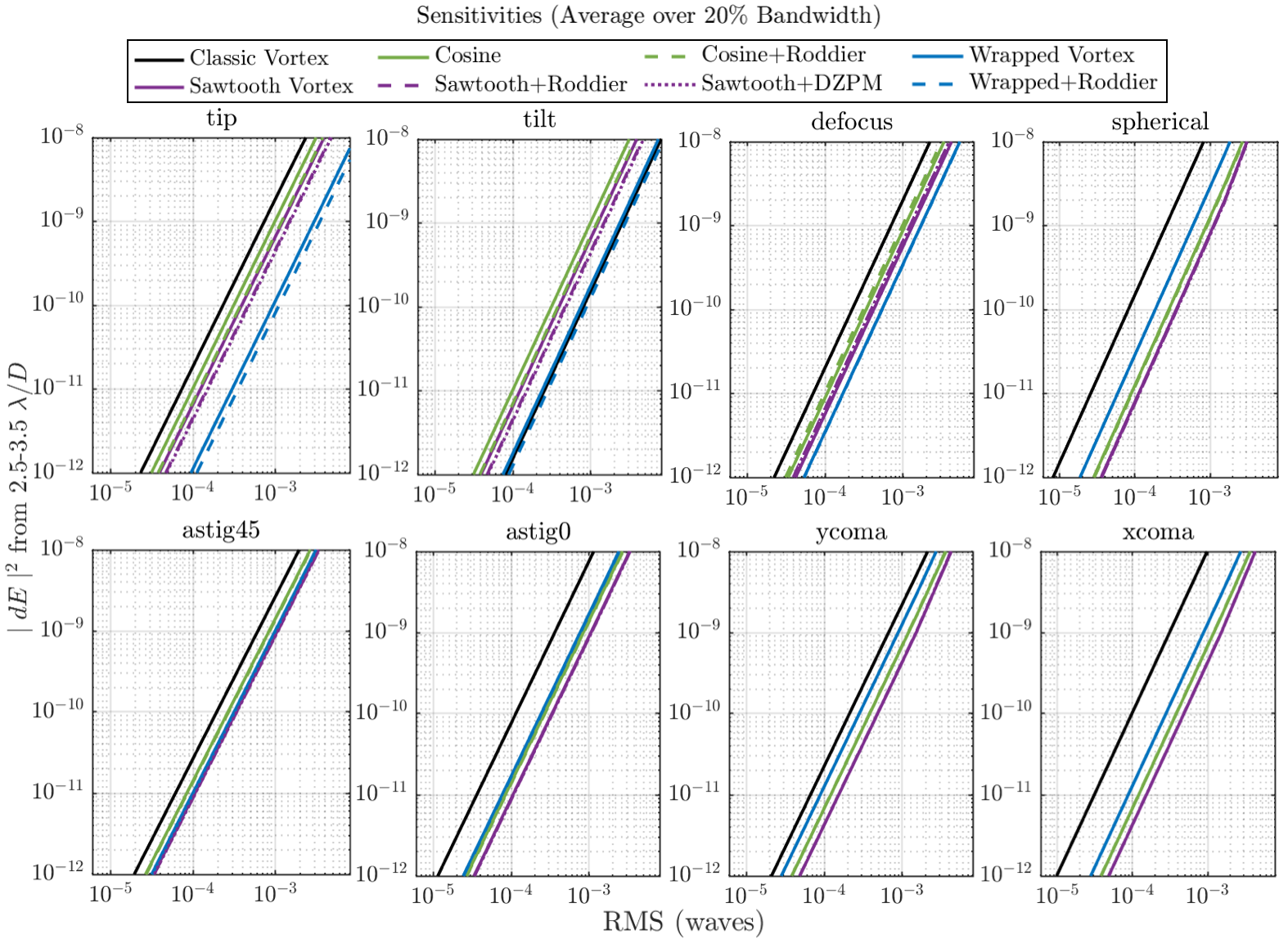}
\end{center}
\caption[fig:sens2] 
%>>>> use \label inside caption to get Fig. number with \ref{}
{ \label{fig:sens2} 
Broadband sensitivities to the lowest order Zernike aberrations (tip, tilt, defocus, spherical, astig, coma). The average change in stellar leakage across 3 ± 0.5 $\lambda_{0}/D$ is shown as a function of root-mean-square (RMS) wavefront error. The solid lines shown here are the four base focal plane masks (classic vortex, sawtooth vortex, wrapped vortex, and cosine). The dashed and dotted lines correspond to hybrid designs incorporating a Roddier phase dimple and dual zone phase dimple, respectively.}
\end{figure} 

A high-performing focal plane mask appropriate for a future space telescope mission would not only demonstrate good starlight suppression across a large bandwidth, but also be insensitive to low-order wavefront aberrations. Figures~\ref{fig:sens1} and~\ref{fig:sens2} show the sensitivity of these proposed coronagraph phase masks to tip/tilt, defocus, spherical, astigmatism and coma Zernike aberrations for monochromatic and broadband light. Figure~\ref{fig:sens1} simulates wavefront error at $\lambda_{0} = 650$ nm, and Figure~\ref{fig:sens2} shows the averaged 20\% broadband result. The stellar leakage is averaged over effective angular separations across 3 $\pm$ 0.5 $\lambda_{0}/D$, and normalized to the peak leakage without the coronagraph masks, as a function of root-mean-square (RMS) wavefront error for each Zernike aberration. These can be directly compared to current capabilities and requirements on wavefront stability.

A vortex coronagraph's sensitivities follow a notional power law as described in Ruane et al. 2018~\cite{Ruane2018}. The sensitivity to each aberration depends on the vortex charge $l$ and the corresponding Zernike polynomial $Z_{n}^{m}(r / a, \theta)$, where $n$ is the radial degree and $m$ is the azimuthal degree and $a$ is the radius of the circular pupil over which the phase is described. The radial and azimuthal degrees of the Zernike polynomials are nonegative integers and with $n \ge \lvert m\rvert \ge 0$ ($m = 0$ for spherical Zernike polynomials). 
The Zernike polynomials may be written as
\begin{equation}
Z_n^m(r / a, \theta)=R_n^{|m|}(r / a)\left\{\begin{array}{ll}
\cos (m \theta) & m \geq 0 \\
\sin (|m| \theta) & m<0
\end{array}, \quad r \leq a,\right.
\end{equation}
where $R_n^m(r / a)$ is the radial Zernike polynomial given by
\begin{equation}
R_n^m(r / a)=\sum_{k=0}^{\frac{n-m}{2}} \frac{(-1)^k(n-k) !}{k !\left(\frac{n+m}{2}-k\right) !\left(\frac{n-m}{2}-k\right) !}(r / a)^{n-2 k}, \quad r / a \leq 1
\end{equation}

For a given charge $l$, the vortex coronagraph filters out aberrations corresponding to Zernike polynomials with $n+m<l$, or in other words, it is weakly sensitive to these aberrations. These are the Zernike aberrations in the ``null space'' for a vortex coronagraph. So for charge 6, this includes tip and tilt, defocus, astigmatism, coma and spherical aberrations. A theoretical vortex coronagraph's sensitivity to the Zernike aberrations in the null space of a charge 6 vortex were reported in Ruane et al. 2018~\cite{Ruane2018} and can be seen in Table~\ref{tab:powerlaws}. In monochromatic light at the design wavelength, all of the scalar vortex designs follow this behavior. However power law fits of the sensitivity orders in Figure~\ref{fig:sens2} reveal that the scalar vortex coronagraphs degrade the out-of-band sensitivity for all the Zernike aberrations in the null space to 2nd order power laws (Table~\ref{tab:powerlaws}). Adding a phase dimple does not restore the notional sensitivity power law properties in broadband light, which we postulate to be due to the still dominant chromatic leakage.

% \begin{table}[!htbp]
%     \begin{center}
%     \includegraphics[height=6cm]{powerlaws.png}
%     \end{center}
%     \caption[tab:powerlaws] 
%     %>>>> use \label inside caption to get Fig. number with \ref{}
%     { \label{tab:powerlaws} 
%     Power laws measured from sensitivity plots in Figures~\ref{fig:sens1} and ~\ref{fig:sens2} for each of the Zernike aberrations in the null space of a charge 6 vortex coronagraph.}
%     \end{table} 

{\renewcommand{\arraystretch}{1.4}
\begin{table} [!htbp]
    \centering
    \resizebox{\columnwidth}{!}{%
    \begin{tabular}{llccc}
    \toprule
        Zernike & $Z^{m}_{n}$ & \begin{tabular}{@{}c@{}}Theoretical charge 6 \\ [-0.5em] vortex coronagraph\end{tabular} & \begin{tabular}{@{}c@{}}Monochromatic SVC \\ [-0.5em] (with and without dimple)\end{tabular} & \begin{tabular}{@{}c@{}}Broadband SVC \\ [-0.5em] (with and without dimple)\end{tabular}\\ 
        \midrule
         tip/tilt & $Z^{\pm1}_{1}$ & 6 & 6 & 2\\ 
         defocus & $Z^{0}_{2}$ & 6 & 6 & 2\\ 
         astigmatism & $Z^{\pm2}_{2}$ & 4 & 4 & 2\\ 
         coma & $Z^{\pm1}_{3}$ & 4 & 4 & 2\\ 
         spherical & $Z^{0}_{4}$ & 4 & 4 & 2\\
         \bottomrule
    \end{tabular}
    }
        \vspace{3 mm} 
    \caption{\label{tab:powerlaws} Zernike Aberration Sensitivities: Power Laws}
\end{table}
}

In monochromatic light, there is not a significant difference in sensitivity between the scalar phase mask topographies. Figures~\ref{fig:sens1} and~\ref{fig:sens2} show that overall there is not a significant impact to the sensitivities to low-order aberrations when a Roddier dimple or DZPM dimple is added. Table~\ref{tab:powerlaws} reports the power laws measured from sensitivity plots in Figures~\ref{fig:sens1} and ~\ref{fig:sens2} for each of the Zernike aberrations in the null space of a charge 6 vortex coronagraph. Table~\ref{tab:powerlaws} shows that in 20\% broadband light, the nominal sensitivity power law is not restored by attenuating the central leakage with a phase dimple.  For the other aberrations: spherical, coma, and astigmatism aberrations, the phase masks with the Roddier dimple or DZPM dimple (dashed or dotted lines) are indistinguishable from the phase masks without them (solid lines). Comparing between the scalar phase mask designs for tip, tilt and defocus, the wrapped vortex has the least sensitivity followed by the sawtooth vortex and then the cosine design. The sawtooth vortex is the most robust to spherical, astigmatism and coma, followed by the cosine design, the wrapped vortex and the classic scalar vortex is the most sensitive. This indicates for the sawtooth design, larger errors may be tolerated on these low order aberrations, which typically dominate the dynamic wavefront error budget.

\subsection{Optimizing the Phase Dimple}
\label{subsec:sweep}

\begin{figure} [b]
\begin{center}
\includegraphics[height=8cm]{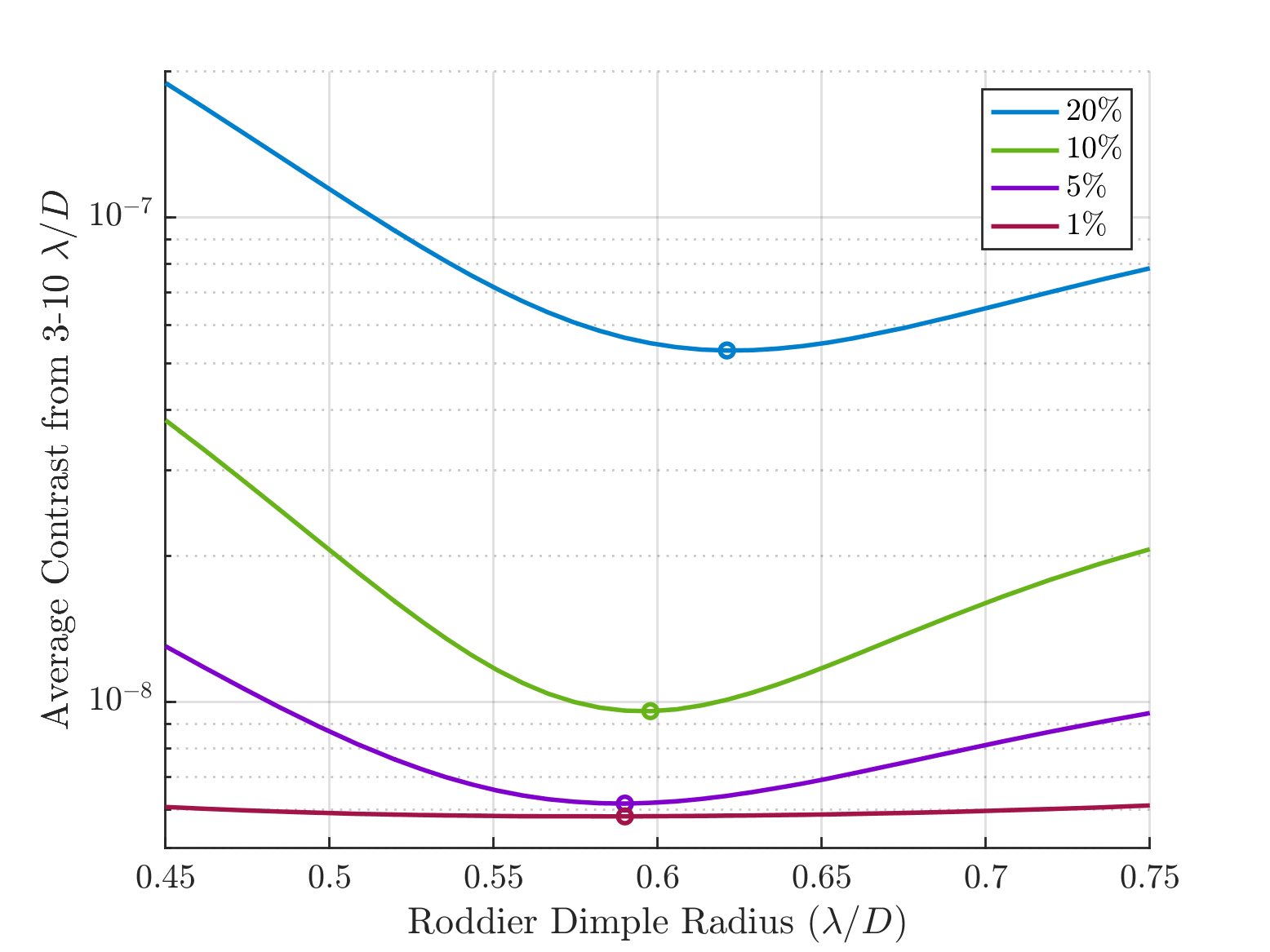}
\end{center}
\caption[fig:rads] 
%>>>> use \label inside caption to get Fig. number with \ref{}
{ \label{fig:rads} 
Sweep of Roddier phase dimple sizes in combination with the sawtooth vortex. Average contrast in final focal plane from 3-10 $\lambda_{0}/D$ shown for 1\%, 5\%, 10\%, and 20\% bandwidths (note 5 wavelengths were sampled across the bandpass to reduce computation). The optimal radius for each bandwidth is marked with a circular marker and values are reported in Table~\ref{tab:radtable}.}
\end{figure}

{\renewcommand{\arraystretch}{1.4}
\begin{table} [!htbp]
    \centering
    \begin{tabular}{cc}
    \toprule
        Bandwidth & Radius ($\lambda/D$)\\ 
        \midrule
         1\% & 0.590\\ 
         5\% & 0.590\\ 
         10\% & 0.600\\ 
         20\% & 0.621\\ 
         \bottomrule
    \end{tabular}
        \vspace{3 mm} 
    \caption{\label{tab:radtable} Optimal radius of Roddier phase dimples for Sawtooth SVC}
\end{table}
}

The next step is to investigate whether the parameters of the Roddier dimple require re-optimization when paired with a sawtooth vortex. Figure~\ref{fig:rads} shows how the average raw contrast from 3-10 $\lambda_{0}/D$ varies with bandwidth and dimple size. This simulation found that for monochromatic light, the contrast is insensitive to a slightly larger or smaller phase dimple. However as the bandwidth increases, the optimal Roddier dimple radius grows slightly larger. The optimal dimple radius values for each of the bandwidths is shown in Table~\ref{tab:radtable}. The DZPM mask is not considered in this optimization since little contrast improvement was observed in the 2D wavefront propagation results (see Table~\ref{tab:contrasts}).

\section{CONCLUSION}
\label{sec:conc}

We investigated the potential benefit of adding radial phase dimples to scalar vortex masks to improve broadband performance, with the goal of addressing 0th order leakage in current scalar vortex designs. A Roddier dimple was superimposed on a sawtooth vortex design, wrapped vortex design, and a cosine design. Additionally a DZPM dimple was superimposed on a sawtooth vortex design. Simulations of 2D wavefront propagation found that adding a Roddier dimple improves the sawtooth vortex average contrast by a factor of 53× and 26× in a 10\% and 20\% bandwidths respectively. With the DZPM dimple, the sawtooth vortex average contrast is improved by the same factor of 53× in 10\% bandwidth, and by a factor of 49× in 20\% bandwidth. Furthermore, for the cosine phase mask, adding a Roddier dimple was found to improve the average contrast by 66× and 46×, in 10\% and 20\%, respectively. Lastly, for the wrapped vortex mask, which already offers improved broadband performance compared to the classic vortex, adding the Roddier dimple was only found to have a moderate impact of roughly a 4× factor of improvement to contrast for both 10\% and 20\%. 

Sensitivity analysis demonstrated that adding the radial phase dimples does not restore the notional sensitivity to low-order wavefront aberrations power law, as expected. Scalar vortex masks, even with suppressed zeroth order leakage, are still significantly more sensitive to aberrations out-of-band and follow a quadratic power law. The sawtooth design has also been identified as more insensitive to most low-order aberrations than the cosine or wrapped vortex designs. This design offers a factor of almost 100× improvement in contrast over the classic vortex, and considering both the sensitivity robustness and contrast performance, it is a likely choice for next steps of manufacturing and testing on a high contrast optical bench.

% Furthermore, for the purpose of selecting a scalar phase mask design to be fabricated, 

These results show that radial phase mask dimples are a promising technique to apply to new scalar vortex designs. A sweep of the Roddier phase dimple parameters indicates a shift in optimal size for different bandwidths. This suggests that it may be essential to consider re-optimizing the parameters when integrating radial features like the Roddier dimple with other scalar phase masks. Although the standalone Roddier and DZPM coronagraphs demonstrate better performance with apodized entrance pupils, these apodizers do not work well with the vortex coronagraph, which prefers circular uniform apertures. Recent experiments with carbon nanotube microdot apodizers in combination with the vortex coronagraph have demonstrated successful elimination of starlight leakage due to segment gap diffraction to the $10^{-8}$ level.\cite{BertrouCantou2023} Similar apodizers could easily be combined with the hybrid designs presented here for application in ground-based telescopes with obstructed apertures.

While adding a central dimple improves simulated performances by up to two orders of magnitude in some cases, the predicted dark regions are still roughly two orders of magnitude away from the ultimate \num{1e-10} level needed for terrestrial planet observations from space. Nevertheless, the performance is certainly more than sufficient for use on the next generation of very large ground-based telescopes. To reach the HWO target contrast value of \num{1e-10} across a 20\% bandwidths, additional achromatization techniques will thus likely be necessary, such as using phase dimples together with achromat plate combinations\cite{Swartzlander2006,Ruane2019} or metasurfaces\cite{konig2023, Palatnick23}.

\subsection*{Disclosure}
This paper is an extension of the work submitted in the SPIE Proceedings: Desai et al. 2023.

\subsection*{Code and Data Availability}
All data in support of the findings of this paper are available within the article or as supplementary material.

\acknowledgments % equivalent to \section*{ACKNOWLEDGMENTS}       
 This work was supported by the NASA ROSES APRA program, grant NM0018F610. Part of this research was carried out at the Jet Propulsion Laboratory, California Institute of Technology, under a contract with the National Aeronautics and Space Administration (80NM00018D0004).

\appendix    % this command starts appendixes

%%%%% References %%%%%

\bibliography{report}   % bibliography data in report.bib
\bibliographystyle{spiejour}   % makes bibtex use spiejour.bst

%%%%% Biographies of authors %%%%%

% \vspace{1ex}
% \noindent Biographies and photographs of the other authors are not available.

% \listoffigures

\end{spacing}
% \end{multicols}

\end{document}